\documentclass[a4paper,twoside]{article}
\newcommand{\affil}[1]{$^{\rm #1}$}
\usepackage[authoryear]{natbib}
\bibpunct{(}{)}{;}{a}{}{,}
\usepackage{graphicx}
\date{} 
\title{\large\bf\flushleft The orbital period of Nova V2540 Ophiuchi (2002)}
\author{\parbox{\textwidth}{\flushleft
\vspace{-0.5cm}
{\it T.~Ak\affil{1}, A.~Retter\affil{2} and A.~Liu\affil{3} \\
\vspace{0.4cm}
{\small \affil{1}\,Istanbul University, Faculty of Science, Dept. of Astronomy 
                   and Space Sciences, 34119, University, Istanbul, Turkey 
                   (tanselak@istanbul.edu.tr)}\\
{\small \affil{2}\,Pennsylvania State University, Department of Astronomy and 
                   Astrophysics, 525 Davey Lab., University Park, PA 16802-6305, 
                   USA (retter@astro.psu.edu)}\\
{\small \affil{3}\,Norcape Observatory, PO Box 300, Exmouth, 6707, Australia 
                   (asliu@onaustralia.com.au)}\\}}}
\begin{document}
\twocolumn[
\begin{changemargin}{.8cm}{.5cm}
\begin{minipage}{.9\textwidth}
\vspace{-1cm}
\maketitle
%
%
\small{\bf Abstract: }We present the results of 26 nights of CCD photometry of the nova 
V2540 Oph (2002) from 2003 and 2004. We find a period of $0.284781$ $\pm$ 0.000006 d 
($6.8347$ $\pm$ 0.0001 h) in the data. Since this period was present in the light curves 
taken in both years, with no apparent change in its value or amplitude, we interpret 
it as the orbital period of the nova binary system. The mass-period relation 
for cataclysmic variables yields a secondary mass of about 0.75 $\pm$ 0.04 $M_{\odot}$.
From maximum magnitude $-$ rate of decline relation, we estimate a maximum absolute 
visual magnitude of $M_{V}$=$-$6.2 $\pm$ 0.4 mag. This value leads to an uncorrected 
distance modulus of $(m-M)$ = 14.7 $\pm$ 0.7. By using the interstellar reddening for 
the location of V2540 Oph, we find a rough estimate for the distance of 5.2 $\pm$ 0.8 kpc. 
We propose that V2540 Oph is either: $\it 1)$ a high inclination cataclysmic variable 
showing a reflection effect of the secondary star, or having a spiral structure in the 
accretion disc, $\it 2)$ a high inclination intermediate polar system, or less 
likely $\it 3)$ a polar. 


\medskip
\medskip
\end{minipage}
\end{changemargin}
]
\small

\section{Introduction}

Nova V2540 Oph ($\alpha_{2000.0}$ = $17^h$ $37^m$ $34.38^s$, 
$\delta_{2000.0}$ = $-$$16^{\circ}$ $23^{'}$ $18.2^{''}$;
\citealp{Katoetal2002}) was independently discovered by Haseda and Nakamura at 
magnitude V=9.0 on 2002 January 24 \citep{Hasedaetal2002}. Examination of 
photographs revealed that the nova reached V=8.9 five days earlier 
\citep{Sekietal2002}. The upper limit on the pre-nova magnitude was estimated 
as 21 mag, indicating a lower limit of about 12.5 for the outburst amplitude 
by adopting an observed maximum magnitude of 8.5 (Kato et al. \citeyear{Katoetal2002}). 
Kato et al. \citeyearpar{Katoetal2002} estimated an upper limit of $M_{V}$=5.7 for 
the nova progenitor. They proposed that the nova should either have a short 
orbital period or a high inclination angle. 

\citet{Retteretal2002a} obtained spectra (400-700 nm) of the nova on 2002 January 
26 and detected strong emission lines of hydrogen and possibly weak P-Cyg profiles. 
They also found Fe II lines, indicating that the object is an Fe II class nova 
caught in the early decline stage. Infrared spectroscopy of V2540 Oph 
\citep{Puetteretal2002} made about 172 days after discovery revealed that the nova 
showed a very rich emission line spectrum with the strongest line at 1.0830 micron 
(He I) and low exitation features of O I and Fe II. The so-called one micron iron 
lines (Fe II) were the strongest lines that have been observed in novae. 

The possibility that V2540 Oph is either a short orbital period nova or a high 
inclination system as suggested by Kato et al. \citeyearpar{Katoetal2002} encouraged 
us to observe the system and to look for photometric periodicities. A preliminary 
report on the detection of a periodicity of about 0.28 d in V2540 Oph was given 
by \citet{Aketal2004} and here we present a compherensive analysis of this finding.

\section{Observations}

V2540 Oph was observed during 16 nights on May and June, 2003, and 10 nights 
on May and July, 2004. The observations span 26 days (167.9 hours in total). Table 1 
presents a summary of the observation schedule. The photometry was carried out with 
a 0.3-m f/6.3 telescope coupled to an ST7 NABG CCD camera. The telescope is located 
in Exmouth, Western Australia, and no filter was used. The exposure times were between 
30 and 60 sec every 120 sec. We estimated differential magnitudes with respect to 
GSC6248-1014 (the comparison star), using GSC6248-1077 as the check star. Their GSC 
magnitudes are 11.30 and 9.96 mag respectively. Differential magnitudes were 
calculated using aperture photometry. The mean GSC magnitude of the comparison star 
was added to the differential magnitudes to give a rough estimate of the visual 
magnitude. The observed magnitudes of V2540 Oph and the comparison star are listed 
in Table 2\footnote{available in electronic form}.

\begin{table}[h]
\begin{center}
\caption[]{The observations timetable.}
\small
\begin{tabular}{lccc}
\hline
Date       & Time of Start & Run Time & Points       \\
           & (HJD-2450000) & (hours)   &  number      \\
\hline
13/5/2003 & 2773.11350    &  2.1     & 60            \\
21/5/2003 & 2781.08614    &  5.8     & 145           \\
22/5/2003 & 2782.11612    &  6.5     & 173           \\
23/5/2003 & 2783.07643    &  7.9     & 210           \\
25/5/2003 & 2785.07620    &  7.4     & 197           \\
26/5/2003 & 2786.07321    &  7.7     & 211           \\
27/5/2003 & 2787.07082    &  7.7     & 202           \\
20/7/2003 & 2840.96941    &  2.7     & 77            \\
23/7/2003 & 2844.02231    &  5.8     & 162           \\
24/7/2003 & 2845.00454    &  6.3     & 173           \\
25/7/2003 & 2845.99641    &  6.5     & 164           \\
26/7/2003 & 2846.98986    &  2.3     & 67            \\
27/7/2003 & 2847.98599    &  6.8     & 185           \\
28/7/2003 & 2848.99353    &  6.5     & 173           \\
29/7/2003 & 2849.99128    &  2.6     & 70            \\
30/7/2003 & 2850.98312    &  6.5     & 191           \\
22/5/2004 & 3148.11717    &  6.8     & 183           \\
25/5/2004 & 3151.13223    &  6.3     & 175           \\
27/5/2004 & 3153.10968    &  6.9     & 193           \\
28/5/2004 & 3154.09891    &  7.2     & 199           \\
29/5/2004 & 3155.08938    &  7.6     & 208           \\
16/6/2004 & 3172.99258    &  8.8     & 135           \\
17/6/2004 & 3174.00195    &  8.3     & 302           \\
18/6/2004 & 3174.99647    &  8.4     & 238           \\
19/6/2004 & 3176.00047    &  8.4     & 239           \\
21/6/2004 & 3178.02243    &  8.1     & 221           \\
\hline
\end{tabular}
\end{center}
\end{table}

\begin{figure}[h]
\begin{center}
\includegraphics[scale=0.265, angle=0]{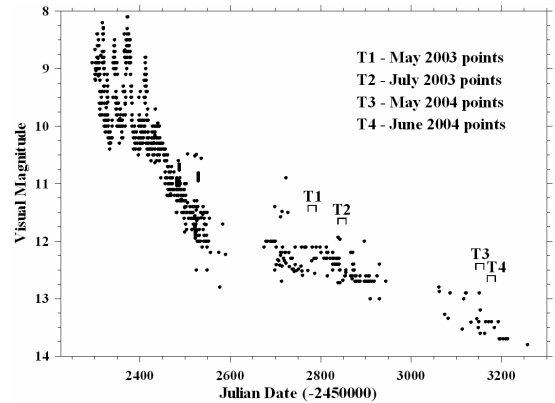}
\caption{\small A 2.65 year light curve of V2540 Oph. Data points are visual 
estimates by amateur astronomers, compiled by AFOEV and AAVSO. The times of our 
observations are marked.}
\end{center}
\end{figure}

\begin{figure}[h]
\begin{center}
\includegraphics[scale=0.375, angle=0]{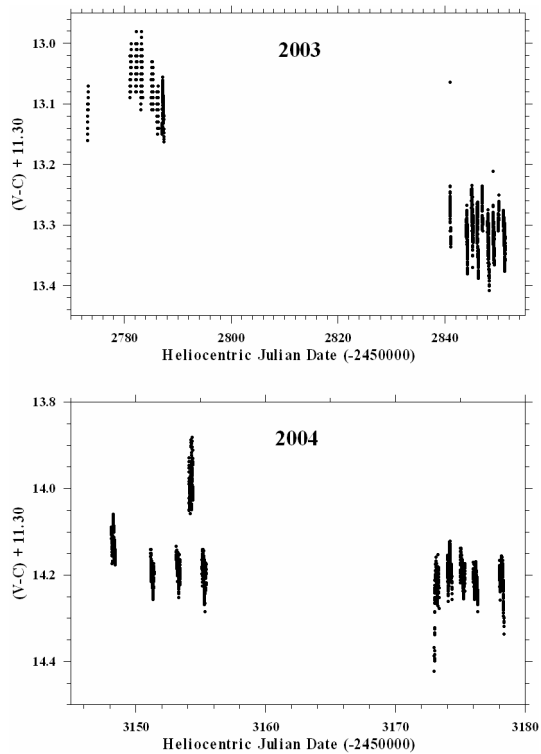}
\caption{\small The light curves of V2540 Oph obtained in the 2003 (upper panel) 
and 2004 (lower panel) observations.}
\end{center}
\end{figure}

\begin{figure}[h]
\begin{center}
\includegraphics[scale=0.375, angle=0]{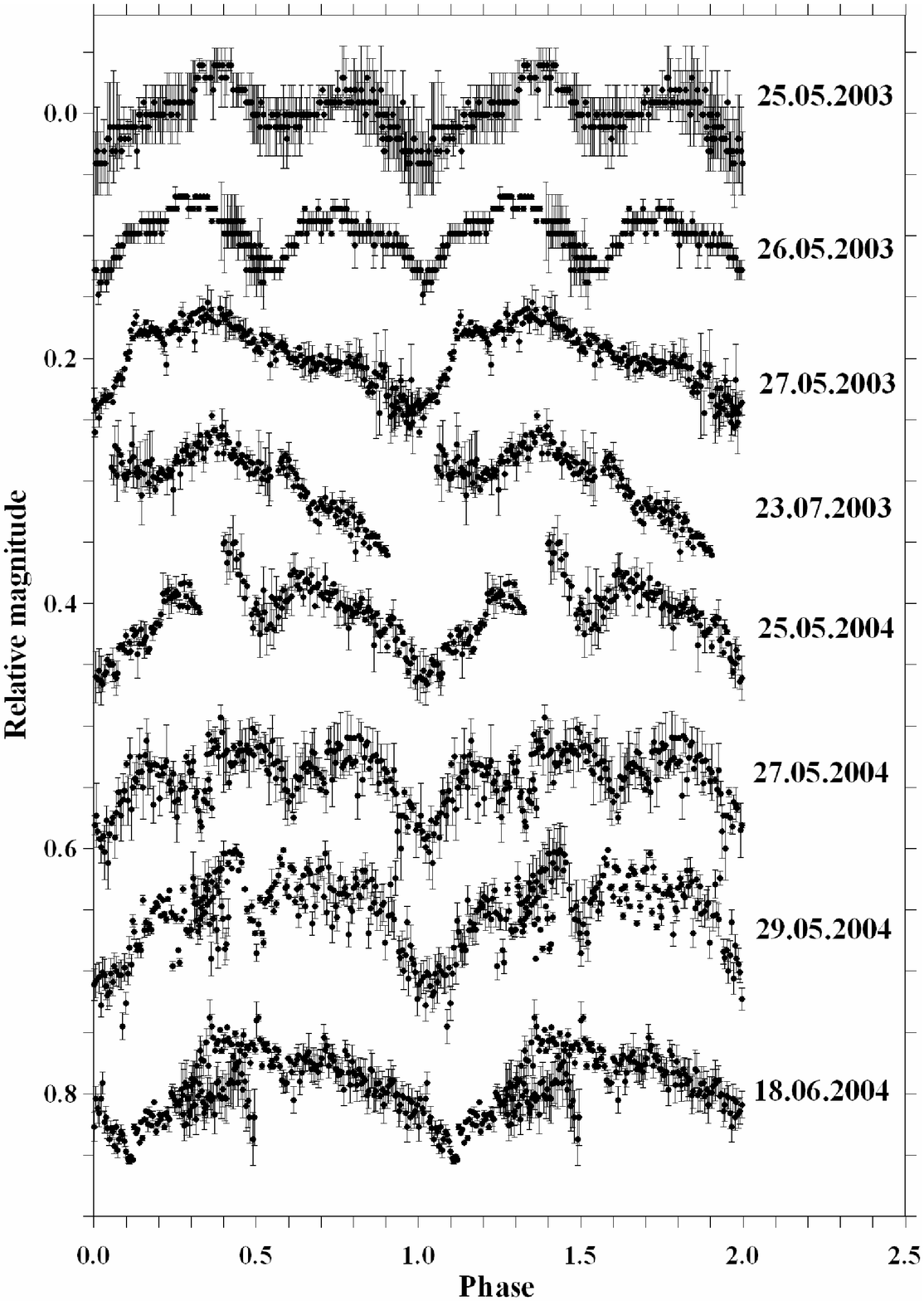}
\caption{\small Samples of orbital light curves in 2003 and 2004.  The orbital phase 
was calculated with the ephemeris given in Section 3.5.}
\end{center}
\end{figure}

Figure 1 displays the visual light curve of the nova from outburst until Sept. 8, 2004. 
The data were taken from the Association Fran\c{c}aise des Observateurs d'Etoiles 
Variables (AFOEV\footnote{http://cdsweb.u-strasbg.fr/afoev/}, Centre de Donnees 
Astronomiques de Strasbourg (CDS)) and the American Association of Variable Star 
Observers (AAVSO\footnote{www.aavso.org}). This data set also includes the 
observations published in IAU Circulars. By combining the data from these 
associations of amateurs, we obtained almost 3300 individual points. Quasi-periodic 
brightness oscillations with a peak-to-peak amplitude of about 1.5 mag were apparent 
during the first 4 months after discovery. These are typical of the transition phase 
that is observed in certain novae. The times of our observations are marked on the graph. 
During the time interval spanned by our observations, the nova declined by about 1.1 mag. 
Kato et al. \citeyearpar{Katoetal2002} indicated that the true maximum of the nova 
could have been missed. However, they also noted that the reported spectrum 
\citep{Retteretal2002a} suggests that the object was caught during an early decay stage. 
The magnitude of the nova was V=8.9 five days before the discovery \citep{Sekietal2002}, 
which is only 0.1 mag fainter than the discovery magnitude, and Kato et al. 
\citeyearpar{Katoetal2002} estimated an upper limit of V=8.5 for the maximum magnitude. 
So, it is possible that the true maximum of the nova was missed. 

We present the light curves of V2540 Oph obtained in 2003 and 2004 separately in 
Figure 2. The jump of about 0.25 mag in the brightness of V2540 Oph in the fourth night 
in 2004 (May 28 2004) is probably real since we can not find a similar feature
in the K-C magnitudes. Here K and C represent unfiltered magnitudes of the comparison star 
and the check star, respectively. The observational errors were estimated from the deviations 
of the K-C magnitudes from the nightly means and are typically about 0.016 and 0.012 mag for 
the 2003 and 2004 observations, respectively. A visual inspection of single runs shows 
a brightness modulation with a period of about 0.3 d in most nights. Samples of orbital 
light curves in 2003 and 2004 are shown in Figure 3.

\section{Data Analysis}

\subsection{The periodogram analysis}

The period analysis was performed using the Data Compensated Discrete Fourier 
Transform (DCDFT, \citealt{FerrazMello1981}, \citealt{Foster1995}), including 
the CLEAN algoritm \citep{Roberstsetal1987}. The DCDFT method is based on 
a least-square regression on two trial functions, sin($f$t) and cos($f$t), and 
a constant. Here $f$ denotes the frequency. In the period analysis, we assume 
that the frequency, say $f_{1}$, that corresponds to the highest peak in the 
power spectrum is real and subtract it from the data. Then, we find the highest 
peak, say $f_{2}$, in the power spectrum of the residuals, subtract $f_{1}$ and 
$f_{2}$ simultaneously from the raw data and calculate a new power spectrum etc. 
until the strongest residual peak is below a given cutoff level. To select the 
peaks, we followed a conservative approach which is similar to the method 
described by \citet{Bregeretal1993} who gave a good criterion for the 
significance of a peak in the power spectrum. In Breger's method, the peaks in 
the power spectrum which are higher than the signal to noise ratio, S/N, of 4.0 
for the amplitude are indicators of real signals. In order to assign a 
confidence level to the power spectra, we calculated the standard error 
($\sigma$) of the power values between the frequencies for which no strong 
peaks appear. We assumed 4$\sigma$ to be the confidence level for the power and 
considered only those peaks of the power spectrum whose power was above this 
level. We then applied the CLEAN algorithm to remove the false peaks until the 
strongest residual peak is below the calculated confidence level. Note that we 
also searched for periodic brightness modulations by Period98 \citep{Sperl1998}, 
which is based on a least-square regression on a trial function, sin($f$t), along 
with a zero point. We found very similar power spectra from both techniques. We 
calculated the error in a frequency from the half width at the half maximum of 
the peak which is a good rough estimator of the uncertainty in a frequency. 
Before applying the power spectrum routines we normalized the data by subtracting 
the mean magnitudes from each night.

\begin{figure}[h]
\begin{center}
\includegraphics[scale=0.265, angle=0]{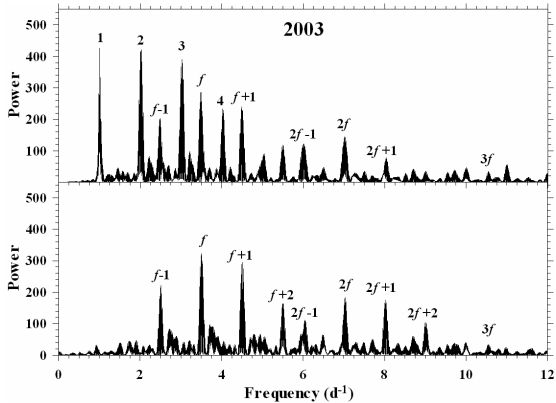}
\caption{\small Upper panel: Power spectrum of the 2003 data after the mean 
magnitudes of the single nights were subtracted from the observations. The peaks at 
the frequencies 1, 2, 3,... $d^{-1}$ originate from the observational gaps. 
Lower panel: The power spectrum of the residuals after subtracting 1 $d^{-1}$ 
aliases. The peaks at the frequencies $f$ = $3.498$ $d^{-1}$, 2$f$, 3$f$ and most 
of their aliases are marked.}
\end{center}
\end{figure}

\subsection{The 2003 light curve}

The power spectrum of the light curve of V2540 Oph obtained in 2003 is shown in the 
upper panel of Figure 4. We removed from the data the nights in which the duration 
of the observations was shorter than 3 hr. The power spectrum is dominated by the 1 $d^{-1}$ 
alias and its harmonics which originate from the observational gaps. Since the 
observations were done from a single site, the 1 $d^{-1}$ alias is very strong, as 
expected. After the 1 $d^{-1}$ alias and its harmonics were subtracted from the data, the 
strongest peak in the power spectrum of the residuals corresponds to the frequency of 
$f$ = 3.498 $\pm$ 0.002 $d^{-1}$. In order to find 
a confidence level for the power, we calculated the standard error ($\sigma$) of 
the power level to be 3.09 between 10-20 $d^{-1}$ after subtracting the harmonics of 
$f$. By considering this standard error as the noise level, we calculated the 
confidence level to be 4$\sigma$=12.36 for the power, as described above.

To further check the reliability of the peak at the frequency 3.498 $d^{-1}$, 
we separately analyzed the data obtained in May and July, 2003. 
In the power spectra of both parts of the data, the same peak ($f$) 
appeared as the strongest one. Thus, we concluded that the peak at the 
frequency 3.498 $d^{-1}$ represents a real periodicity in the 2003 data.

To search for additional signals, $f$, its harmonics, and the 1 $d^{-1}$ aliases were 
subtracted from the data. The power levels of the peaks in the power spectrum of the 
residuals are generally lower than the confidence level. The peaks indicating 
frequencies shorter than 3 $d^{-1}$ are artefacts of the removed frequencies and 
correspond to periodicities that are longer than the typical interval of observations 
in each night. Thus, we can not consider them as real signals. In summary, we concluded 
that there are no additional signals in the 2003 light curve.

\subsection{The 2004 light curve}

The power spectrum of the 2004 light curve of V2540 Oph is shown 
in the upper panel of Figure 5. The strongest peak in the power spectrum corresponds 
to the frequency of $f$ = 3.51 $\pm$ 0.01 $d^{-1}$, which is consistent 
with that found in the 2003 data within errors. The 1 $d^{-1}$ alias 
and its harmonics which originate from the observational gaps are very strong, 
as well. We calculated the standard error ($\sigma$) of the 
power level to be 2.20 between 10-20 $d^{-1}$ after subtracting the harmonics of $f$. 
By considering this standard error as the noise level, we calculated the 
confidence level to be 4$\sigma$=8.80 for the power.

\begin{figure}[h]
\begin{center}
\includegraphics[scale=0.265, angle=0]{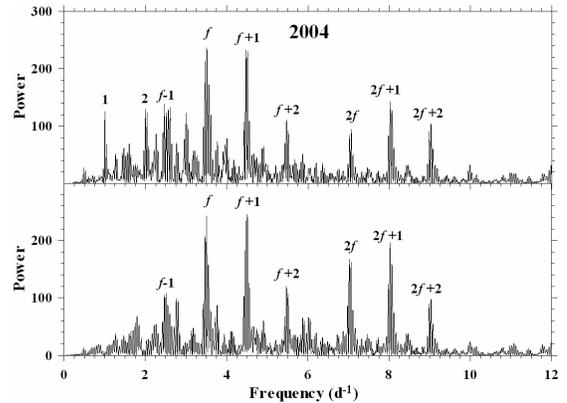}
\caption{\small Upper panel: Power spectrum of the 2004 data after 
the mean magnitudes of each night were subtracted from the 
observations. Lower panel: The power spectrum of the residuals after 
subtracting 1 $d^{-1}$ aliases. The peak at the frequency 
$f$ = $3.51$ $d^{-1}$ is marked with its first overtone and their 
aliases.}
\end{center}
\end{figure}

To further check the significance of the peak at the frequency 3.51 $d^{-1}$, 
we divided the data into two distinct parts (May and June, 2004). 
In the power spectra of both parts of the observations, the same peak ($f$) 
appeared as the strongest one. Thus, we concluded that the peak at the 
frequency 3.51 $d^{-1}$ indicates a real periodicity in the 2004 data.

In summary, the frequencies of the signals found in 2003 and 2004 data 
are consistent with each other within errors.

\begin{figure}[h]
\begin{center}
\includegraphics[scale=0.265, angle=0]{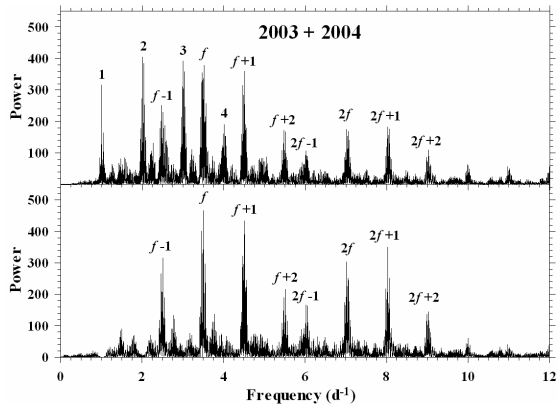}
\caption{\small Upper panel: Power spectrum of the combined data. 
Lower panel: The power spectrum of the residuals after subtracting 
1 $d^{-1}$ aliases. The peak at the frequency $f$ = $3.51147$ $d^{-1}$ is marked 
as well as its first overtone and their aliases.}
\end{center}
\end{figure}

\subsection{The power spectrum of the combined data}

Analysis of the combined light curve observed in the two consequtive seasons can
give a more precise value of the periodicity. To combine the data obtained 
in the two seasons, the cumulative error during the observing gap should be smaller 
than one cycle. The observed frequency and its error in 2003 data were given 
above as $f$ = 3.498 $\pm$ 0.002 $d^{-1}$. Thus, after $\sim$1750 cycles 
(df/f$\sim$5.72$\times$$10^{-4}$) the cumulative error is one cycle and 
we can not combine data obtained in two separate seasons. Fortunately, the gap 
between the observations from the two seasons is $\sim$300 days, or 
$\sim$1050 cycles. This calculation shows that the data from the two seasons 
can be combined. 

Figure 6 is a plot of the power spectrum of all data except runs shorter than 3 h. 
The highest peak in the power spectrum corresponds to the frequency of 
$f$ = 3.51147 $\pm$ 0.00008 $d^{-1}$ ($0.284781$ $\pm$ 0.000006 d). Strong peaks at 
the 1 $d^{-1}$ aliases and its harmonics are also prominent. 

It should be noted that we could not find a statistically significant 
peak in the frequency range 20-100 $d^{-1}$ in the power spectra of the 
2003, 2004 and the combined light curves.

\subsection{Structure of the periodicity}

In Figure 7 we show the light curve of V2540 Oph folded on the $0.284781$ d period. 
We omitted from the data the nights shorter than one full cycle. The points are the 
average magnitude value in each of the 40 equal bins that cover the 0$-$1 phase interval. 
The bars are 1$\sigma$ uncertainties in the average values. The full 
amplitude of the mean variation is 0.060 $\pm$ 0.004 and 0.061 $\pm$ 0.002 mag for 
the 2003 and 2004 observations, respectively. The amplitudes were derived by dividing 
the folded data into 10 equal intervals and by measuring the difference between 
extrema. 

The best fitted ephemeris of the periodicity is:
\[
T_{min}(HJD) = 2453151.316(6) + 0.284534(8) E \\ 
\]
Since the error in this value is larger than the result from the periodogram analysis, 
we adopt instead the value $0.284781$ $\pm$ 0.000006 d found from the periodogram 
analysis of the combined light curve for the period.

A noteworthy feature seen in Figure 7 is the dip in the mean light curve at phase 0.5. 
The dip is present in both seasons and in the combined data. Its reliability was checked 
by subtracting a pure sinusoidal with a peak-to-peak amplitude of 0.060 mag from the 
folded and binned data, resulting in a distribution of the points around zero level. 
This method gave an eclipse-like feature at phase 0.5 with a full amplitude of 
about 0.02 mag.

\begin{figure}[h]
\begin{center}
\includegraphics[scale=0.37, angle=0]{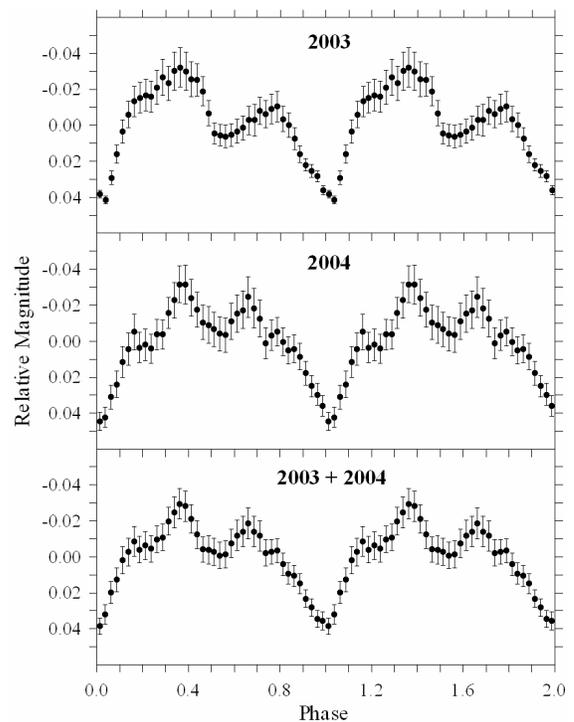}
\caption{\small The light curves of V2540 Oph obtained in 2003 (upper panel) and 
2004 (middle panel), folded on the 0.284781 d period and binned into 40 equal bins. 
The lowest panel shows the combined light curve folded on the same period. 
Note the dip at phase 0.5.}
\end{center}
\end{figure}

\subsection{The long-term light curve of the nova}

We estimated a few parameters of the nova using the visual magnitude estimates by 
amateur astronomers that were made available by AAVSO and AFOEV (Section 2; Figure 1). 
As the typical uncertainty of a visual estimate is about $\pm$0.3 mag, we have 
determined a mean light curve by taking 0.1 d averages of the individual points. 
This led to an averaged light curve containing 705 points. 

We found a decay rate of 0.011 $\pm$ 0.002 mag/d during the 
first 296 days (JD 2452294-2452590) of the long-term light curve of V2540 Oph by 
fitting polinomials. However, at first glance, we see that the light curve can be  
divided into 5 linear parts that have different decay rates (see Figure 1). We 
calculated decay rates by linear fits to these parts of the light curve. The fading of 
the system was relatively fast (0.039 $\pm$ 0.004 mag/d) during the first 47 days 
(JD 2452294-2452341) of the light curve. This decay rate is compatible with 
0.033 mag/d measured by Kato et al. \citeyearpar{Katoetal2002} for the same time 
interval. However, during the next 99 days, the fading slowed down with a very low 
decay rate of 0.003 $\pm$ 0.001 mag/d. In the following part of the light curve, 
the system faded with a decay rate (0.015 $\pm$ 0.002 mag/d) faster than the previous part
for about 150 days. Then the light curve showed another slow fading phase with 
0.002 $\pm$ 0.001 mag/d lasting about 270 days. The decay rate of the last part of the 
light curve was 0.005 $\pm$ 0.001 mag/d which makes this part of the light curve 
steeper than the previuos one. Similar patterns can be seen in the light curves 
of nova V443 Sct \citep{Anupamaetal1992} and nova V4745 Sgr \citep{Csaketal2005}.

By using the maximum visual magnitude of 8.5 (Kato et al. \citeyear{Katoetal2002}) we 
measured $t_{2}$ = 158 $\pm$ 10 d and $t_{3}$ = 213 $\pm$ 15 d from the long-term 
light curve which makes V2540 Oph a very slow nova according 
to the classification given in Table 5.4 of \citet{Warner1995}. The $t_{2}$ and $t_{3}$ 
parameters roughly obey the relation $t_{3}\approx 2.75t_{2}^{0.88}$ given by 
\citet{Warner1995}. Note the uncertainty in these values due to the possibility that the 
maximum was missed and because of the oscillations during the transition phase 
(Section 1; Figure 1). 

We calculated the visual absolute magnitude in maximum with four maximum magnitude versus 
rate of decline relations by \citet{DownesandDuerbeck2000}, \citet{DellaValleandLivio1995},
\citet{Capacciolietal1989} and \citet{Cohen1985} using $t_{2}$ = 158 $\pm$ 10 d. Our $t_{2}$ 
measurement resulted in $-$5.71, $-$6.86, $-$6.79, $-$5.40 mag, respectively. We adopted the 
formal average of these values, which is $M_{V}$=$-$6.2 $\pm$ 0.4 mag. This value yields an 
uncorrected distance modulus of $(m-M)$ = 14.7 $\pm$ 0.7 using the maximum visual magnitude 
of 8.5 (Kato et al. \citeyear{Katoetal2002}). The galactic reddening map by  
\citet{Schlegeletal1998}\footnote{http://nedwww.ipac.caltech.edu/} yields an upper limit 
of $E(B-V)$ = 0.452 mag for the position of V2540 Oph. This yields an extinction of about
$A_{V}$ = 1.5 mag. Consequently, we find a distance between 4.4 and 6.0 kpc. 

\citet{KissandThomson2000} proposed that the quasi periods of the oscillations seen 
in nova light curves have a similar time scale of $t_{2}$. V2540 Oph does not obey 
this relation since its $t_{2}$ parameter is much longer than the quasi period of 
the post-maximum oscillations which is about 32 $\pm$ 1 d (see below). It should be noted 
that $t_{2}$ and the related parameters may be changed if the maximum visual magnitude of 
the nova was brighter than 8.5, say 8.0-8.2 mag. In this case, $t_{2}$ is measured 
as about 50 days, which is closer to the quasi-periodicity of the oscillations.

Another striking feature in the long-term light curve of V2540 Oph is the post-maximum 
oscillations with a full amplitude of about 1.5 mag superimposed on the fading phase during
JD 2452294-2452440. The periodogram analysis gives a characteristic recurrence time of 
32 $\pm$ 1 d. However, it is evident that the maxima were not repeating periodically. 
We determined epochs of maxima by fitting low-order polinomials to the selected parts of 
the light curve. An inspection of the epochs shows that the time interval between successive 
maxima tends to increase monotonically in time. Similar trends were observed for GK Per, 
DK Lac \citep{Bianchinietal1992} and V4745 Sgr \citep{Csaketal2005}.

\section{Discussion}

We identified one periodicity in the light curve of V2540 Oph about 16 
and 28 months after its outburst. We suggest that the periodicity found in 
this study, P $\sim$ $6.83$ h, is the orbital period of the nova binary system. 
Such a period is typical of orbital periods in novae and cataclysmic variables 
\citep{Warner2002}. The fact that the amplitude of the variation was very similar 
during the two years of observations (Section 3.5) despite a fading of 1.1 mag suggests 
that the dominant light source is the varying component, which is likely the accretion 
structure - the accretion disc or the accretion stream. 

Using the orbital period of the system, we can obtain a rough 
mass estimate for the secondary star of the system. \citet{SmithandDhillon1998} 
derived a mass-period relation for the secondary stars in CVs using a sample 
of reliable mass estimates. From their Equation (9) we obtain a mass of 
0.75 $\pm$ 0.04 $M_{\odot}$ for the secondary star in V2540 Oph. The error in 
the mass was calculated from the standard deviation of the fit parameters. 
Using a mean white dwarf mass of 0.85 $\pm$ 0.05 $M_{\odot}$ given for 
classical novae \citep{SmithandDhillon1998}, we find a mass ratio of 
$M_{2}$/$M_{1}$ = 0.9 $\pm$ 0.1. In the following we discuss several explanations 
for the observed variation.

\subsection{A superhump periodicity?}

The periodicity found in the light curve of V2540 Oph may be interpreted as 
a superhump period. Superhumps are quasi-periodic modulations that appear in the 
light curves of some cataclysmic variables (CVs), and in nova-like systems they 
are called permanent superhumps 
\citep{Pattersonetal1997,Patterson1999,Patterson2001,RetterandNaylor2000,Retteretal2002b}. 
The superhump period is a few per cent longer or shorter than the orbital period.
The commonly accepted interpretation of the superhump phenomenon is that the light 
modulations result from the beat periodicity between the orbital period and 
the precession of the accretion disc around the white dwarf 
of the underlying binary system. However, our suggestion of the orbital period is 
based mainly on the following facts: $\it 1)$ This period was present in the light 
curves taken in 2003 and 2004, with no apparent change in its value, $\it 2)$ We did 
not observe a considerable amplitude difference of this brightness modulation between 
the 2003 and 2004 observations, and $\it 3)$ The shape of the light curve does not support 
a superhump period, since the mean shape of superhumps is typically an asymmetric 
sinusoid \citep[e.g.,][]{Pattersonetal1997,Retteretal1997,Retteretal2003}. 
Thus, we conclude that it is unlikely that the observed periodicity is a superhump.

\subsection{Ellipsoidal effect of the secondary star}

Another possible explanation of the variation in the light curve is the ellipsoidal effect 
of the companion star \citep[e.g.,][]{Retteretal1999}. Using Equation (2.102) 
of \citet{Warner1995} with the period of 6.83 h, we find that the visual absolute magnitude 
of the red dwarf is only about $M_{V(sec)}$$\approx$$+$7.4. An inspection of the long-term 
visual observations shows that the average visual magnitude of the nova was 12.3 and 13.4 
during our 2003 and 2004 observations, respectively. If we assume that the distance of 
V2540 Oph is 4.4-6.0 kpc, we find an absolute magnitude between $-1.3$ and $-3.1$ for the 
system using $A_{V}$ = 1.5 mag (see Section 3.6). These absolute magnitudes show that 
the contribution of the secondary star to the total light of the system is less 
than 1 per cent, lower than observed. Thus, the light variation of V2540 Oph can probably 
not be attributed to the ellipsoidal effect of the companion star since the secondary 
star is too faint.

\subsection{Irradiation effect and the polar interpretation}

Retter et al. \citeyearpar{Retteretal1999} investigated the plausibility of the 
brightness modulation in the nova DN Gem originating from the irradiation of the 
secondary star using the model described in \citet{Somersetal1996} and \citet{Ioannouetal1999}. 
They concluded that an irradiated secondary star seems a likely source of the 
brightness modulations. These are similar to those seen in the light curve of 
V2540 Oph. 

\citet{WoudtandWarner2003} noted that one of the following requirements must be 
fulfilled for the large amplitude orbital modulations seen in the light curve of a recent 
nova in which the accretion disc does not dominate the luminosity of the system : 
$\it 1)$ The disc is foreshortened but the irradiated secondary is seen (high inclination 
angle), $\it 2)$ The disc has small dimensions (a short orbital period), $\it 3)$ No disc 
(the system is a polar). For V2540 Oph, the disc, if present, must be large since the 
system has a relatively long orbital period for a cataclysmic variable. Since the orbital 
period implies a hot secondary star due to its relatively high mass ($\sim$0.75 $M_{\odot}$), 
this star can produce a significant irradiation effect. The folded light curve of V2540 Oph 
with a dip at phase 0.5 (Figure 7) resembles  that of the polar V834 Cen 
\citep{Cropperetal1986} and the nova and polar candidate V2214 Oph \citep{Baptistaetal1993}. 
However, the possibility that V2540 Oph is a polar seems weak as the orbital period distribution 
of polars concentrates around orbital periods shorter than 5 h \citep{Warner1995}. Thus, 
using above requirements we can say that V2540 Oph is possibly a high inclination system with 
an irradiated secondary. If the brightness modulation seen in nova V2540 Oph originates 
from the irradiation of the secondary star, the dip at phase 0.5 may be an eclipse of the 
light reflected from the secondary star. 

\citet{Katoetal2004} found a bump like feature at phase 0.6-0.7 and a dip-like feature at 
phase 0.2-0.4 in the light curve of the eclipsing transient nova V1494 Aql. These features are 
similar to those seen in the light curve of V2540 Oph. They suspected that the structure 
in the accretion disc fixed in the binary rotational frame may be responsible for these 
features. \citet{Hachisuetal2004} showed that the out-of-eclipse features of V1494 Aql 
can be explained by spiral shocks on the accretion disk in the late phase of the nova 
outburst. A similar model can explain the out-of-eclipse features observed in V2540 Oph 
as well. 

Although we can not unambiguously reject the possibility that the system is a polar, 
we can simply conclude that V2540 Oph is likely a high inclination system either showing an 
irradiation effect or having a spiral structure in its accretion disc. Interestingly, 
Kato et al. \citeyearpar{Katoetal2002} proposed that the nova should have either a short 
orbital period or a high inclination angle.

\subsection{The transition phase and the intermediate polar model}

The quasi-periodic brightness oscillations with a characteristic recurrence 
time of $\sim$32 d seen during the first four months of the long-term light curve 
of V2540 Oph (Figure 1) show that the nova was caught in the transition phase. We found a
monotonic increase of the time interval between successive maxima in time (Section 3.6). 
Such trends were observed in GK Per, DK Lac (Bianchini et al. \citeyear{Bianchinietal1992}) 
and V4745 Sgr \citep{Csaketal2005}. The optical light curve of a classical nova is 
typically characterized by a smooth decline. However, certain novae display a deep 
minimum in the light curve while others have slow oscillations, during the 
so-called '{\it transition phase}'. The minimum is understood by the formation of a dust 
envelope around the binary system. 

It should be noted that the term '{\it transition phase}' generally applies to 
a phenomenon between optical maximum and final fading stage, typically a few 
weeks-months after maximum. The oscillations observed in the long-term light curve 
of V2540 Oph occurred in a very early decay stage, soon after discovery. 
Thus, they may not represent the same phenomenon as typical transition phase 
oscillations. However, if the true maximum of the nova (see Section 2) was missed, 
the oscillations seen in the light curve would be the transition phase oscillations. 

Several models have been suggested for the transition phase, such as stellar oscillations 
of the hot white dwarf, oscillations caused by the wind, formation of dust blobs that move 
in and out of line of sight to the nova, dwarf nova outbursts, mini nova-outbursts and 
oscillations of the common envelope \citep{BodeandEvans1989,Leibowitz1993,Warner1995,Csaketal2005}.
\citet{Shaviv2001} argued that super-Eddington winds can be a natural explanation for 
the oscillatory behaviour during the transition phase of some novae.
\citet{Retter2002} suggested another solution for the transition phase problem and proposed 
a possible connection between the transition phase and intermediate polars (IPs)
\citep[for a review on IPs see e.g.,][]{Hellier1999}.

The IP model for V2540 Oph can not be ruled out, although we could not find a short-term 
spin period for the white dwarf in the light curve of V2540 Oph. In the transient nova 
V1494 Aql, no short-term period was found in the optical, but a $\sim$40 min period was detected
in the X-ray \citep{Drakeetal2003}. A short-term X-ray period of $\sim$22 min was also found 
in the transient nova V4743 Sgr \citep{Nessetal2003}. Thus, to confirm or refute the 
suggestion that novae with transition phase are intermediate polars, further X-ray and 
optical observations of novae are required. 

\citet{Schmidtobreicketal2005} argued that some tremendous outburst 
nova candidates are likely to originate from rather low mass transfer rate systems, 
i.e. dwarf novae. However, they also expect that these systems have orbital periods 
shorter than classical novae. Thus, considering its orbital period suggested in our study, 
we believe that the possibility that the origin of V2540 Oph is a dwarf nova is weak.

As a concluding remark, we can say that the nova V2540 Oph may be: $\it 1)$ a high 
inclination cataclysmic variable with an irradiated secondary star or a spiral structure 
in its accretion disc, $\it 2)$ a high inclination intermediate polar system, or $\it 3)$ a polar. 
It should be noted that the last possibility is fairly weak considering the orbital period 
distribution of polars.

\section*{Acknowledgments}

The authors would like to thank the anonymous referee for useful comments that helped 
improving an early version of the paper. 

This work was partially supported by a postdoctoral fellowship from Penn State University. 
Part of this work was also supported by the Research Fund of the University of Istanbul, 
Project Numbers: BYP-724/24062005 and BYP-738/07072005. We acknowledge the observers of 
the AAVSO and AFOEV who made the observations that comprise the long-term light curve 
of the nova V2540 Oph used in this study.

\end{document}